\documentclass[11pt,a4paper]{article}

\usepackage{amssymb}
\usepackage[dvips]{graphicx}
\usepackage{bm}

\usepackage{mathrsfs}
\usepackage{cite}

\begin{document}

\title{Exact $\beta$-functions for ${\cal N}=1$ supersymmetric theories finite in the lowest loops}

\author{
K.V.Stepanyantz\\
{\small{\em Moscow State University,}}\\
{\small{\em Faculty of Physics, Department of Theoretical Physics,}}\\
{\small{\em 119991, Moscow, Russia}}}

\maketitle

\begin{abstract}
We consider a one-loop finite ${\cal N}=1$ supersymmetric theory in such a renormalization scheme that the first $L$ contributions to the gauge $\beta$-function and the first $(L-1)$ contributions to the anomalous dimension of the matter superfields and to the Yukawa $\beta$-function vanish. It is demonstrated that in this case the NSVZ equation and the exact equation for the Yukawa $\beta$-function in the first nontrivial order are valid for an arbitrary renormalization prescription respecting the above assumption. This implies that under this assumption the $(L+1)$-loop contribution to the gauge $\beta$-function and the $L$-loop contribution to the Yukawa $\beta$-function are always expressed in terms of the $L$-loop contribution to the anomalous dimension of the matter superfields. This statement generalizes the result of Grisaru, Milewski, and Zanon that for a theory finite in $L$ loops the $(L+1)$-loop contribution to the $\beta$-function also vanishes. In particular, it gives a simple explanation why their result is valid although the NSVZ equation does not hold in an arbitrary subtraction scheme.
\end{abstract}

\unitlength=1cm

\section{Introduction}
\hspace*{\parindent}

The Yukawa and gauge $\beta$-functions in ${\cal N}=1$ supersymmetric theories are related to the anomalous dimension of the matter superfields. The former exact equation follows from the non-renormalization theorem for superpotential \cite{Grisaru:1979wc}. Really, due to the absence of divergent quantum corrections to superpotential, it is possible (although not necessary) to choose a subtraction scheme in which

\begin{equation}\label{Lambda_Renormalization}
\lambda^{ijk} = \lambda_0^{mnp} (Z_\phi^{1/2})_m{}^i (Z_\phi^{1/2})_n{}^j (Z_\phi^{1/2})_p{}^k,
\end{equation}

\noindent
where $\lambda_0^{ijk}$ and $\lambda^{ijk}$ are the bare and renormalized Yukawa couplings, respectively, and $(Z_\phi)_i{}^j$ is the renormalization constant for the matter superfields. In this scheme the Yukawa $\beta$-function is related to the matter superfield anomalous dimension exactly in all loops,

\begin{equation}\label{Yukawa_Beta_Minimal}
(\beta_\lambda)^{ijk}(\alpha,\lambda) = \frac{1}{2} \Big((\gamma_\phi)_m{}^i \lambda^{mjk} + (\gamma_\phi)_m{}^j \lambda^{imk} + (\gamma_\phi)_m{}^k \lambda^{ijm}\Big)
= \frac{3}{2}\, (\gamma_\phi)_m{}^{(i}(\alpha,\lambda) \lambda^{jk)m},
\end{equation}

\noindent
where the indices inside the round brackets are symmetrized,

\begin{equation}
T^{(ijk)} \equiv \frac{1}{3!}\Big(T^{ijk} +\mbox{permutations of $ijk$}\Big).
\end{equation}

Note that there are some other non-renormalization theorems in ${\cal N}=1$ supersymmetric theories. For example, the triple gauge-ghost vertices are finite in all loops \cite{Stepanyantz:2016gtk}. (Recently this statement has been confirmed by an explicit calculation of the two-loop matter contribution to these vertices \cite{Kuzmichev:2021yjo}.) However, the most interesting relation, which can be considered as a non-renormalization theorem, is the exact NSVZ $\beta$-function \cite{Novikov:1983uc,Jones:1983ip,Novikov:1985rd,Shifman:1986zi}. It is an all-loop equation relating the $\beta$-function of ${\cal N}=1$ supersymmetric gauge theories to the anomalous dimension of the matter superfields. For a theory with a simple gauge group $G$ and chiral matter superfields in the representation $R$ (with the generators $T^A$) it is written as

\begin{equation}\label{NSVZ_Beta1}
\frac{\beta(\alpha,\lambda)}{\alpha^2} = - \frac{3 C_2 - T(R) + C(R)_i{}^j (\gamma_\phi)_j{}^i(\alpha,\lambda)/r}{2\pi(1- C_2\alpha/2\pi)},
\end{equation}

\noindent
where $\alpha$ is the renormalized gauge coupling constant and $r = \mbox{dim}\, G$  is the dimension of the gauge group. (In our notation the bare gauge coupling constant is denoted by $\alpha_0$.) The group Casimirs are defined as

\begin{equation}
\mbox{tr}(T^A T^B) \equiv T(R)\, \delta^{AB};\qquad  (T^A T^A)_i{}^j \equiv C(R)_i{}^j; \qquad f^{ACD} f^{BCD} \equiv C_2\delta^{AB}
\end{equation}

\noindent
under the assumption that the generators of the fundamental representation $t^A$ are normalized by the condition $\mbox{tr}(t^A t^B) = \delta^{AB}/2$. For theories with extended supersymmetry the NSVZ equation (\ref{NSVZ_Beta1}) gives the non-renormalization theorems (first derived in \cite{Grisaru:1982zh,Mandelstam:1982cb,Brink:1982pd,Howe:1983sr}, see also \cite{Buchbinder:1997ib}) provided a quantization procedure respects at least ${\cal N}=2$ supersymmetry \cite{Shifman:1999mv,Buchbinder:2014wra,Buchbinder:2015eva}. NSVZ analogs were written for theories with softly broken supersymmetry \cite{Hisano:1997ua,Jack:1997pa,Avdeev:1997vx} and for the Adler $D$-function in ${\cal N}=1$ SQCD \cite{Shifman:2014cya,Shifman:2015doa}. Eq. (\ref{NSVZ_Beta1}) can also be applied for investigating finite ${\cal N}=1$ supersymmetric theories, see, e.g., \cite{Heinemeyer:2019vbc} and references therein, and even finite theories with softly broken supersymmetry \cite{Kazakov:1995cy,Kazakov:1997nf}. Really, if the NSVZ equation is valid, then the $L$-loop finiteness immediately leads to the vanishing of the $(L+1)$-loop $\beta$-function. However, it is well known that the NSVZ $\beta$-function is scheme-dependent and is valid only in certain (NSVZ) subtraction schemes, which do not include $\overline{\mbox{DR}}$ \cite{Jack:1996vg,Jack:1996cn,Jack:1998uj,Harlander:2006xq} (see also the review \cite{Mihaila:2013wma}) and MOM \cite{Kataev:2013csa,Kataev:2014gxa} schemes. All-loop prescriptions giving an NSVZ scheme have been constructed in \cite{Kataev:2013eta} for ${\cal N}=1$ SQED and in \cite{Stepanyantz:2020uke} for a general renormalizable ${\cal N}=1$ supersymmetric gauge theory with a simple gauge group. Both constructions are based on the underlying results of \cite{Stepanyantz:2011jy,Stepanyantz:2014ima} for ${\cal N}=1$ SQED and of \cite{Stepanyantz:2016gtk,Stepanyantz:2019ihw,Stepanyantz:2019lfm,Stepanyantz:2020uke} for the non-Abelian case. For all theories mentioned above the NSVZ scheme is given by the HD+MSL prescription.\footnote{In the Abelian case another all-loop NSVZ renormalization presription is the on-shell scheme \cite{Kataev:2019olb}.} This means that a theory is regularized by Higher covariant Derivatives \cite{Slavnov:1971aw,Slavnov:1972sq} in a supersymmetric version \cite{Krivoshchekov:1978xg,West:1985jx}\footnote{This regularization also includes insertion of the Pauli--Villars determinants for regularizing one-loop divergences \cite{Slavnov:1977zf}, see \cite{Aleshin:2016yvj,Kazantsev:2017fdc} for the corresponding supersymmetric construction. Also it is very essential that the ratios of the Pauli--Villars masses to the dimensionful cut-off parameter $\Lambda$ inside higher derivative terms should not depend on couplings.} and divergences are removed by Minimal Subtractions of Logarithms. In other words, renormalization constants should contain only powers of $\ln\Lambda/\mu$ without any finite terms. Various NSVZ-like relations are also valid in the HD+MSL-scheme \cite{Nartsev:2016mvn,Kataev:2017qvk}.\footnote{For theories with softly broken supersymmetry this has yet been proved only for the Abelian case.} Again the proof is based on the underlying results of \cite{Nartsev:2016nym} and \cite{Shifman:2014cya,Shifman:2015doa}. Note that the HD+MSL scheme is not in general unique, because minimal subtractions of logarithms can supplement different versions of the higher covariant derivative regularization. Therefore, this renormalization prescription can give a continuous family of the NSVZ (or NSVZ-like) schemes. (It is known that NSVZ schemes constitute a continuous set in both Abelian \cite{Goriachuk:2018cac} and non-Abelian \cite{Goriachuk_Conference,Goriachuk:2020wyn} case.)

The fact that the HD+MSL prescription gives the NSVZ schemes has been confirmed by a large number of explicit calculations in such orders where the scheme dependence is essential, see, e.g., \cite{Stepanyantz:2012zz,Shakhmanov:2017soc,Kazantsev:2018nbl,Kuzmichev:2019ywn,Aleshin:2020gec}. Note that  in technical aspects these calculations are rather similar to those made for theories with higher derivatives (see, e.g., \cite{BezerradeMello:2016bjn,Gama:2017ets,Gama:2020pte}), but the structure of divergences in higher derivative theories and in usual theories for which higher derivative terms are introduced only for a regularization is certainly quite different.

Possibly, the all-loop prescription giving an NSVZ scheme could help to construct a renormalization scheme in which one-loop finite ${\cal N}=1$ supersymmetric theories are finite in all orders. Its existence follows from the results of \cite{Kazakov:1986bs,Ermushev:1986cu,Lucchesi:1987he,Lucchesi:1987ef}, but an explicit all-loop prescription has not yet been found. In this paper we will discuss the scheme dependence of the renormalization group functions (RGFs) if some lowest terms in the $\beta$-functions and in the anomalous dimension of the matter superfields vanish. In this case the scheme dependence can have some special features, see, e.g., \cite{Ryttov:2020vsz}. Concerning the (gauge and Yukawa) $\beta$-functions, we will demonstrate that in the first nontrivial order the exact equations for them are valid for an arbitrary renormalization prescription which respects the finiteness in the previous orders. For theories in which the one-loop anomalous dimension and the two-loop gauge $\beta$-function vanish such a feature of the three-loop gauge $\beta$-function has been established by an explicit calculation made in \cite{Kazantsev:2020kfl}. Here we generalize this result and discuss its consequences.

The paper is organized as follows. In Sect. \ref{Section_Three_Loops}, following Ref. \cite{Kazantsev:2020kfl}, we recall how the three-loop $\beta$-function is related to the two-loop anomalous dimension for one-loop finite theories. General equations describing the scheme dependence of RGFs are written in Sect. \ref{Section_Scheme_Dependence}. Using these equations in Sect. \ref{Section_L=2} we rederive the result of Ref. \cite{Kazantsev:2020kfl} starting from the relations valid in the HD+MSL scheme. Finite renormalizations which do not break two-loop finiteness are constructed in Sect. \ref{Section_Finiteness_For_L=3}. The result is generalized in Sect.  \ref{Section_Main_Result}, where we obtain finite renormalizations which do not break the $L$-loop finiteness. They are used for proving the relations which express the $(L+1)$-loop gauge and the $L$-loop Yukawa $\beta$-functions in terms of the $L$-loop anomalous dimension of the matter superfields for an arbitrary renormalization prescription.

\section{The three-loop result}
\hspace*{\parindent}\label{Section_Three_Loops}

As a starting point we consider a one-loop finite ${\cal N}=1$ supersymmetric gauge theory. For such a theory the gauge group Casimirs and the coupling constants should satisfy the conditions

\begin{equation}\label{1Loop_Finiteness}
T(R) = 3C_2; \qquad \lambda^*_{imn} \lambda^{jmn} = 4\pi\alpha C(R)_i{}^j.
\end{equation}

\noindent
The vanishing of the one-loop $\beta$-function follows from the left equation (which is an analog of the Banks--Zaks condition \cite{Banks:1981nn}), and the vanishing of the one-loop anomalous dimension of matter superfields is expressed by the right equation. According to \cite{Parkes:1984dh}, the second equation in (\ref{1Loop_Finiteness}) also leads to the vanishing of the two-loop $\beta$-function and even to the vanishing of the two-loop anomalous dimension in the $\overline{\mbox{DR}}$-scheme. The latter statement was also derived from arguments based on anomalies \cite{Jones:1983vk,Jones:1984cx}. Note that in the two-loop approximation the $\beta$-function is scheme independent, so that the former statement is true for an arbitrary renormalization prescription. Evidently, it also follows from the NSVZ $\beta$-function (\ref{NSVZ_Beta1}). As for the latter statement, although the constraints (\ref{1Loop_Finiteness}) lead to the vanishing of the two-loop anomalous dimension in the $\overline{\mbox{DR}}$-scheme \cite{Parkes:1984dh,Jones:1983vk,Jones:1984cx}, for a general renormalization prescription it is not true \cite{Parkes:1985hh}. For a general renormalization prescription which does not break Eq. (\ref{Lambda_Renormalization}) the two-loop anomalous dimension of a theory satisfying Eq. (\ref{1Loop_Finiteness}) is given by the expression

\begin{eqnarray}\label{Gamma_Finite}
&&\hspace*{-7mm} (\gamma_\phi)_i{}^j(\alpha,\lambda)\Big|_{\mbox{\scriptsize Eq. (\ref{1Loop_Finiteness})}} = -\frac{3\alpha^2}{2\pi^2}C_2 C(R)_i{}^j\Big(\ln \frac{a_{\varphi}}{a} -b_{11}+b_{12}\Big)
- \frac{\alpha}{4\pi^2} \Big(\frac{1}{\pi} \lambda^*_{imn} \lambda^{jml} C(R)_l{}^n \nonumber\\
&&\hspace*{-7mm} + 2\alpha \left[C(R)^2\right]_i{}^j\Big) \Big(A-B -2g_{12} +2g_{11}\Big) + O\Big(\alpha^3,\alpha^2\lambda^2,\alpha\lambda^4,\lambda^6\Big)
\end{eqnarray}

\noindent
derived in \cite{Kazantsev:2020kfl} using the higher covariant derivative regularization. Here the parameters $a=M/\Lambda$ and $a_\varphi = M_\varphi/\Lambda$ are the ratios of Pauli--Villars masses to the cut-off parameter in the higher derivative term. Two other regularization parameters

\begin{equation}\label{AB_Definitions}
A=\int\limits_0^\infty dx \ln x\, \frac{d}{dx}\frac{1}{R(x)};\qquad B = \int\limits_0^\infty dx \ln x\, \frac{d}{dx}\frac{1}{F^2(x)}
\end{equation}

\noindent
are finite constants, which are determined by the higher-derivative regulators $R(x)$ and $F(x)$ in the gauge and matter parts of the action, respectively, see \cite{Kazantsev:2020kfl} for details. The dependence on a regularization is supplemented by the dependence on a renormalization prescription, which is expressed by the presence of the finite constants $b_{11}$, $b_{12}$, $g_{11}$, and $g_{12}$. They appear due to an arbitrariness of choosing finite parts of the renormalization constants and fix a subtraction scheme in the considered approximation,

\begin{eqnarray}\label{Z_Alpha2}
&&\hspace*{-2mm} \frac{1}{\alpha} - \frac{1}{\alpha_0} = -\frac{3}{2\pi}C_2\Big(\ln \frac{\Lambda}{\mu} + b_{11}\Big) + \frac{1}{2\pi} T(R) \Big(\ln \frac{\Lambda}{\mu} + b_{12}\Big) + O(\alpha,\lambda^2);\\
\label{Z_Phi1}
&&\hspace*{-2mm} (Z_\phi)_i{}^j = \delta_i{}^j + \frac{\alpha}{\pi} C(R)_i{}^j \Big(\ln\frac{\Lambda}{\mu}+g_{11}\Big) - \frac{1}{4\pi^2} \lambda^*_{imn}\lambda^{jmn} \Big(\ln\frac{\Lambda}{\mu} + g_{12} \Big)
+ O(\alpha^2,\alpha\lambda^2,\lambda^4).\qquad
\end{eqnarray}

\noindent
Note that in Ref. \cite{Kazantsev:2020kfl} the renormalization of the Yukawa couplings was made according to Eq. (\ref{Lambda_Renormalization}), so that no new finite constants appeared in the corresponding equation.

The most popular $\overline{\mbox{DR}}$ scheme corresponds to

\begin{equation}\label{DR_Constants}
g_{11} = -\frac{1}{2} -\frac{A}{2};\qquad g_{12} = -\frac{1}{2} -\frac{B}{2};\qquad b_{11} = \ln a_\varphi;\qquad b_{12} = \ln a.
\end{equation}

\noindent
(We consider two renormalization schemes as equivalent if they give the same expressions for the renormalized Green functions as well as the same expressions for RGFs defined in terms of the renormalized couplings.) In the HD+MSL scheme all finite constants vanish, so that $g_{11} = g_{12} = b_{11} = b_{12} = 0$.

According to \cite{Kazantsev:2020kfl}, the three-loop $\beta$-function for theories which satisfy Eq. (\ref{1Loop_Finiteness}) is

\begin{eqnarray}\label{Beta_Finite}
&& \frac{\beta(\alpha,\lambda)}{\alpha^2}\bigg|_{\mbox{\scriptsize Eq. (\ref{1Loop_Finiteness})}} = \frac{3\alpha^2}{4\pi^3 r} C_2\, \mbox{tr}\left[C(R)^2\right] \Big(\ln \frac{a_\varphi}{a} + b_{12} - b_{11}\Big) + \frac{\alpha}{8\pi^3 r} \Big(\frac{1}{\pi} C(R)_j{}^i C(R)_l{}^n  \qquad\nonumber\\
&& \times \lambda^*_{imn} \lambda^{jml} +2\alpha\,\mbox{tr}\left[C(R)^3\right]\Big) \Big(A-B -2g_{12}+2g_{11}\Big) + O\Big(\alpha^3,\alpha^2\lambda^2,\alpha\lambda^4,\lambda^6\Big). \vphantom{\frac{1}{\pi^2}}
\end{eqnarray}

\noindent
Again, in general, it does not vanish, but for arbitrary values of finite constants $b_i$ and $g_i$ satisfies the equation

\begin{equation}\label{3L_NSVZ}
\frac{\beta(\alpha,\lambda)}{\alpha^2} = - \frac{1}{2\pi r} C(R)_i{}^j (\gamma_{\phi})_j{}^i(\alpha,\lambda)  + O(\alpha^3,\alpha^2\lambda^2,\alpha\lambda^4,\lambda^6),
\end{equation}

\noindent
which is a consequence of the NSVZ equation (\ref{NSVZ_Beta1}) in the considered approximation. From this equation we immediately conclude that if a theory under consideration is finite in the two-loop approximation, then its three-loop $\beta$-function also vanishes in a complete agreement with the result of \cite{Grisaru:1985tc}. Note that {\it a priory} the use of the NSVZ $\beta$-function for proving this fact is impossible, because Eq. (\ref{NSVZ_Beta1}) is not valid for a general renormalization prescription. However, in what follows we will explain why and how the result of \cite{Grisaru:1985tc} can easily be derived from the NSVZ equation in all loops.

\section{Scheme dependence of renormalization group functions}
\hspace*{\parindent}\label{Section_Scheme_Dependence}

First, let us, for the further convenience, make the formal replacement

\begin{equation}\label{Replacement}
\alpha \to g\alpha;\qquad \lambda^{ijk} \to g^{1/2} \lambda^{ijk};\qquad \lambda^*_{ijk} \to g^{1/2} \lambda^*_{ijk},
\end{equation}

\noindent
where $g$ is a real constant, so that

\begin{eqnarray}
&& \beta(\alpha,\lambda) \equiv \frac{d\alpha}{d\ln\mu}\bigg|_{\alpha_0,\lambda_0 = \mbox{\scriptsize const}} \to\ \sum\limits_{n=1}^\infty \beta_n(\alpha,\lambda)\, g^{n+1};\nonumber\\
&& (\gamma_\phi)_i{}^j(\alpha,\lambda) \equiv \frac{d(\ln Z_\phi)_i{}^j}{d\ln\mu}\bigg|_{\alpha_0,\lambda_0 = \mbox{\scriptsize const}} \to \sum\limits_{n=1}^\infty (\gamma_{\phi,n})_i{}^j(\alpha,\lambda)\, g^{n};\nonumber\\
&& (\beta_\lambda)^{ijk}(\alpha,\lambda) \equiv \frac{d\lambda^{ijk}}{d\ln\mu}\bigg|_{\alpha_0,\lambda_0 = \mbox{\scriptsize const}} \to\ \sum\limits_{n=1}^\infty (\beta_{\lambda,n})^{ijk}(\alpha,\lambda)\, g^{n+1/2}.\qquad
\end{eqnarray}

\noindent
The index of summation $n$ corresponds to the number of loops for a contribution to a certain renormalization group function. Note that here we will not assume that the renormalizations of the Yukawa couplings and of the matter superfields are related by Eq. (\ref{Lambda_Renormalization}). This implies that, in general, the equation

\begin{equation}\label{L_Yukawa_Beta_Minimal}
(\beta_{\lambda,L})^{ijk}(\alpha,\lambda) = \frac{3}{2}\, (\gamma_{\phi,L})_m{}^{(i}(\alpha,\lambda) \lambda^{jk)m},
\end{equation}

\noindent
which follows from Eq. (\ref{Yukawa_Beta_Minimal}), can be broken in higher loops. (As we will demonstrate below, this can occur only for theories which are not finite in the previous loops).

Now, let us consider a theory which satisfies the constraints (\ref{1Loop_Finiteness}). According to \cite{Kazakov:1986bs,Ermushev:1986cu,Lucchesi:1987he,Lucchesi:1987ef}, there exists a renormalization prescription for which this theory is finite in all orders. For example, a scheme in which a one-loop finite theory is finite in the three-loop approximation has explicitly been constructed in Ref. \cite{Jack:1996qq}. However, at present there is no all-loop prescription how to obtain this scheme, so that it has to be tuned in each order. Let us assume that a subtraction scheme is tuned in such a way that the gauge $\beta$-function does not vanish only in the $(L+1)$-loop approximation, while the first nonvanishing contributions to the anomalous dimension of the matter superfields and to the Yukawa $\beta$-function appear only in the $L$-th loop,

\begin{eqnarray}\label{Assumption}
&& \beta_{n}(\alpha,\lambda) = 0,\qquad\qquad\ \, n=1,\ldots,L;\qquad \vphantom{\Big(}\nonumber\\
&& (\beta_{\lambda,n})^{ijk}(\alpha,\lambda)=0,\qquad n=1,\ldots, L-1;\qquad \vphantom{\Big(}\nonumber\\
&& (\gamma_{\phi,n})_i{}^j(\alpha,\lambda) = 0,\qquad\ \, n=1,\ldots, L-1.\qquad \vphantom{\Big(}
\end{eqnarray}

\noindent
We would like to prove that under the assumption (\ref{Assumption}) for an arbitrary renormalization prescription the Yukawa and gauge $\beta$-functions satisfy Eq. (\ref{L_Yukawa_Beta_Minimal}) and the equation

\begin{equation}\label{L_NSVZ}
\frac{\beta_{L+1}(\alpha,\lambda)}{\alpha^2} = - \frac{1}{2\pi r} C(R)_i{}^j (\gamma_{\phi,L})_j{}^i(\alpha,\lambda),
\end{equation}

\noindent
respectively. Certainly, Eq. (\ref{L_NSVZ}) immediately follows from Eqs. (\ref{NSVZ_Beta1}) and (\ref{Assumption}). Similarly, Eq. (\ref{L_Yukawa_Beta_Minimal}) is obtained from Eq. (\ref{Lambda_Renormalization}).
However, Eqs. (\ref{Lambda_Renormalization}) and (\ref{NSVZ_Beta1}) do not hold for a general renormalization prescription. Therefore, the above statement is not trivial. To prove it, we will use the fact that at least one  NSVZ scheme satisfying Eq. (\ref{Lambda_Renormalization}) exists and is given by the HD+MSL prescription \cite{Stepanyantz:2016gtk,Stepanyantz:2019ihw,Stepanyantz:2020uke}. According to \cite{Vladimirov:1979my}, the other renormalization prescriptions can be obtained from this scheme by finite renormalizations of the form

\begin{equation}\label{Finite_Renormalizations}
\alpha \to \alpha'(\alpha,\lambda);\qquad \lambda \to \lambda'(\alpha,\lambda);\qquad
(Z_\phi)_i{}^j \to (Z'_\phi)_i{}^j(\alpha',\lambda',\Lambda/\mu) = z_i{}^k(\alpha,\lambda) (Z_\phi)_k{}^j(\alpha,\lambda,\Lambda/\mu).
\end{equation}

\noindent
We will always assume that these transformations respect finiteness in the previous orders, so that Eq. (\ref{Assumption}) will also hold after them. In the lowest approximations they are written as

\begin{equation}\label{Finite_Renormalizations_Explicit}
\frac{1}{\alpha'} = \frac{1}{\alpha} + g\, b_0 + O(\alpha,\lambda^2);\qquad \lambda'{}^{ijk} = \lambda^{ijk} + O(\alpha\lambda,\lambda^3);\qquad z_i{}^j = \delta_i^j + O(\alpha,\lambda^2),
\end{equation}

\noindent
where $b_0$ is a finite constant. Note that the form of the finite renormalizations is chosen in such a way that it is consistent with a structure of quantum corrections. In particular, from the above equations we see that the change of $\alpha$ does not contain $\alpha\lambda^2$ terms, and the first $\lambda$-dependent contributions are proportional to $\alpha^2\lambda^2$. Under the considered finite renormalization RGFs that we are interested in change as \cite{Vladimirov:1975mx,Vladimirov:1979ib}

\begin{eqnarray}\label{Finite_Renormalization_Beta}
&&\hspace*{-7mm} \beta'(\alpha',\lambda') = \frac{\partial\alpha'}{\partial\alpha} \beta(\alpha,\lambda)
+ g^{1/2} \frac{\partial\alpha'}{\partial\lambda^{ijk}} (\beta_\lambda)^{ijk}(\alpha,\lambda)
+ g^{1/2} \frac{\partial\alpha'}{\partial\lambda^*_{ijk}} (\beta_\lambda^*)_{ijk}(\alpha,\lambda);\\
\label{Finite_Renormalization_Gamma}
&&\hspace*{-7mm} (\gamma_\phi')_i{}^j(\alpha',\lambda') =
(\gamma_\phi)_i{}^j(\alpha,\lambda) + g^{-1} \frac{\partial\ln z_i{}^j}{\partial\alpha}\, \beta(\alpha,\lambda)\nonumber\\
&& \qquad\qquad\qquad\qquad\ \,
+ g^{-1/2} \frac{\partial\ln z_i{}^j}{\partial\lambda^{mnp}}\, (\beta_\lambda)^{mnp}(\alpha,\lambda)
+ g^{-1/2} \frac{\partial\ln z_i{}^j}{\partial\lambda^*_{mnp}}\, (\beta_\lambda^*)_{mnp}(\alpha,\lambda);\qquad\\
\label{Finite_Renormalization_Beta_Lambda}
&&\hspace*{-7mm} (\beta_\lambda')^{ijk}(\alpha',\lambda') = g^{-1/2}\frac{\partial\lambda'{}^{ijk}}{\partial\alpha} \beta(\alpha,\lambda)
+ \frac{\partial\lambda'{}^{ijk}}{\partial\lambda^{mnp}} (\beta_\lambda)^{mnp}(\alpha,\lambda)
+ \frac{\partial\lambda'{}^{ijk}}{\partial\lambda^*_{mnp}} (\beta_\lambda^*)_{mnp}(\alpha,\lambda).\qquad
\end{eqnarray}

\noindent
These equations will be used below for proving that for an arbitrary $L$ Eqs. (\ref{L_Yukawa_Beta_Minimal}) and (\ref{L_NSVZ}) follow from Eq. (\ref{Assumption}). However, it is expedient to start with the particular case $L=2$, which is considered in the next section.

\section{Explicit calculation for $L=2$}
\hspace*{\parindent}\label{Section_L=2}

According to Eq. (\ref{Assumption}), in the particular case $L=2$ we assume that the anomalous dimension of the matter superfields and the Yukawa $\beta$-function vanish in the one-loop approximation, while the gauge $\beta$-function vanishes in the two-loop approximation. These coefficients can be extracted with the help of the replacement (\ref{Replacement}), after which

\begin{eqnarray}\label{Gamma_Lowest}
&& (\gamma_\phi)_i{}^j(\alpha,\lambda) = g (\gamma_{\phi,1})_i{}^j(\alpha,\lambda) + g^2 (\gamma_{\phi,2})_i{}^j(\alpha,\lambda) + O(g^3);\qquad\vphantom{\Big(}\\
\label{Beta_Lowest}
&& \beta(\alpha,\lambda) = g^2 \beta_1(\alpha) + g^3 \beta_2(\alpha,\lambda) + g^4 \beta_3(\alpha,\lambda) + O(g^5); \qquad \vphantom{\Big(}\\
\label{Beta_Lambda_Lowest}
&& (\beta_\lambda)^{ijk}(\alpha,\lambda) = g^{3/2} (\beta_{\lambda,1})^{ijk}(\alpha,\lambda) + g^{5/2} (\beta_{\lambda,2})^{ijk}(\alpha,\lambda) + O(g^{7/2}). \qquad \vphantom{\Big(}
\end{eqnarray}

\noindent
(Note that writing the second equation we took into account that the one-loop $\beta$-function is independent of the Yukawa couplings.) After imposing the condition (\ref{1Loop_Finiteness}) the above assumption is evidently valid for an arbitrary renormalization prescription,

\begin{equation}\label{Assumptions_Lowest}
(\gamma_{\phi,1})_i{}^j(\alpha,\lambda)\Big|_{\mbox{\scriptsize Eq.} (\ref{1Loop_Finiteness})} = 0; \quad\ (\beta_{\lambda,1})^{ijk}(\alpha,\lambda)\Big|_{\mbox{\scriptsize Eq.} (\ref{1Loop_Finiteness})} = 0;\quad\  \beta_1(\alpha)\Big|_{\mbox{\scriptsize Eq.} (\ref{1Loop_Finiteness})} = 0;\quad\ \beta_2(\alpha,\lambda)\Big|_{\mbox{\scriptsize Eq.} (\ref{1Loop_Finiteness})} = 0.
\end{equation}

\noindent
Now, let us suppose that RGFs (\ref{Gamma_Lowest}) --- (\ref{Beta_Lambda_Lowest}) are written in a certain scheme in which Eqs. (\ref{Lambda_Renormalization}) and (\ref{NSVZ_Beta1}) are satisfied. For example, it is so in the HD+MSL scheme. (For this renormalization prescription the NSVZ equation is satisfied according to \cite{Stepanyantz:2020uke}, while Eq. (\ref{Lambda_Renormalization}) evidently holds, because only $\ln\Lambda/\mu$ can appear in both its sides.) In the HD+MSL scheme the lowest coefficients of RGFs satisfy the equations

\begin{eqnarray}\label{Beta23}
&& \beta_2(\alpha,\lambda) = -\frac{\alpha^2}{2\pi r} C(R)_i{}^j (\gamma_{\phi,1})_j{}^i(\alpha,\lambda);\nonumber\\
&& \beta_3(\alpha,\lambda) = -\frac{\alpha^2}{2\pi r} C(R)_i{}^j (\gamma_{\phi,2})_j{}^i(\alpha,\lambda) -\frac{\alpha^3}{4\pi^2 r} C_2 C(R)_i{}^j (\gamma_{\phi,1})_j{}^i(\alpha,\lambda);\qquad\nonumber\\
&& (\beta_{\lambda,1})^{ijk}(\alpha,\lambda) = \frac{3}{2}\, (\gamma_{\phi,1})_m{}^{(i}(\alpha,\lambda)\, \lambda^{jk)m};\nonumber\\
&& (\beta_{\lambda,2})^{ijk}(\alpha,\lambda) = \frac{3}{2}\, (\gamma_{\phi,2})_m{}^{(i}(\alpha,\lambda)\, \lambda^{jk)m}.
\end{eqnarray}

\noindent
Let us perform a finite renormalization

\begin{eqnarray}\label{Finite_Renormalization_Original}
&& g\,\alpha' = g\,\alpha + g^2 \Delta_1\alpha + g^3 \Delta_2\alpha + O(g^4);\qquad\vphantom{\Big(}\nonumber\\
&& g^{1/2}\lambda'{}^{ijk} = g^{1/2} \lambda^{ijk} + g^{3/2} \Delta_1\lambda^{ijk} + O(g^{5/2});\qquad\vphantom{\Big(}\nonumber\\
&& z_i{}^j = \delta_i^j + g\,\Delta_1 z_i{}^j + O(g^2).\vphantom{\Big(}
\end{eqnarray}

\noindent
The new anomalous dimension for one-loop finite theories in the first non-trivial order can be obtained from Eq. (\ref{Finite_Renormalization_Gamma}). Calculating it one needs to take into account that the finiteness conditions analogous to Eq. (\ref{1Loop_Finiteness}) should now be imposed on the new couplings $\alpha'$ and $\lambda'$,

\begin{equation}\label{1Loop_Finiteness_Prime}
T(R) = 3C_2; \qquad \lambda'{}^*_{imn} \lambda'{}^{jmn} = 4\pi\alpha' C(R)_i{}^j.
\end{equation}

\noindent
Note that the terms in the expression (\ref{Finite_Renormalization_Gamma}) containing $\beta$ and $\beta_\lambda$ are of the order $O(g^3)$ due to Eq. (\ref{1Loop_Finiteness_Prime}). This implies that in the lowest nontrivial approximation they can be omitted. Moreover, in order to use Eq. (\ref{Assumptions_Lowest}), we should express the couplings $\alpha$ and $\lambda$ in terms of $\alpha'$ and $\lambda'$ and expand the result as a Taylor series in $\alpha'-\alpha$ and $\lambda'-\lambda$. After this, from Eq. (\ref{Finite_Renormalization_Gamma}) we obtain

\begin{eqnarray}\label{New_Gamma}
&&\hspace*{-5mm} (\gamma'_\phi)_i{}^j(\alpha',\lambda')\Big|_{\mbox{\scriptsize Eq.} (\ref{1Loop_Finiteness_Prime})} = g^2 \Big(-\frac{\partial (\gamma_{\phi,1})_i{}^j}{\partial\alpha}\, \Delta_1\alpha - \frac{\partial (\gamma_{\phi,1})_i{}^j}{\partial\lambda^{mnp}}\, \Delta_1\lambda^{mnp} - \frac{\partial (\gamma_{\phi,1})_i{}^j}{\partial\lambda^*_{mnp}}\, \Delta_1\lambda^*_{mnp} \nonumber\\
&&\hspace*{-5mm} + (\gamma_{\phi,2})_i{}^j\Big)\bigg|_{\alpha\to \alpha';\ \lambda \to \lambda';\ \mbox{\scriptsize Eq.} (\ref{1Loop_Finiteness_Prime})} + O(g^3),
\end{eqnarray}

\noindent
where the subscript `$\alpha\to \alpha'$, $\lambda\to\lambda'$' means that the original arguments $\alpha$ and $\lambda$ of the functions present in the right hand side of Eq. (\ref{New_Gamma}) should be formally replaced by  $\alpha'$ and $\lambda'$. The scheme-independent one-loop contribution to the anomalous dimension entering Eq. (\ref{New_Gamma}) is given by the expression

\begin{equation}\label{1Loop_Gamma}
(\gamma_{\phi,1})_i{}^j = - \frac{\alpha}{\pi}C(R)_i{}^j+\frac{1}{4\pi^2}\lambda^*_{imn}\lambda^{jmn} \equiv \frac{1}{8\pi^2} P_i{}^j(\alpha,\lambda),
\end{equation}

\noindent
where (following Ref. \cite{Jack:1996vg}) we introduced the notation

\begin{equation}\label{P_Definition}
P_i{}^j(\alpha,\lambda) \equiv 2\lambda^*_{imn} \lambda^{jmn} - 8\pi\alpha C(R)_i{}^j.
\end{equation}

To specify the form of finite renormalizations, we present them as a sum of relevant tensor structures containing the coupling constants and group Casimirs. The degrees of coupling constants should be chosen in such a way that after the replacement (\ref{Replacement}) the degrees of $g$ coincide with the ones in Eq. (\ref{Finite_Renormalization_Original}). Various products of group Casimirs should be consistent with the structure of possible quantum corrections in the corresponding order of the perturbation theory. Then the lowest terms in Eq. (\ref{Finite_Renormalization_Original}) can be written in the form

\begin{eqnarray}\label{Finite_Renormalization_Explicit}
&&\hspace*{-5mm} \Delta_1\alpha = \frac{\alpha^2}{2\pi}\Big(3 C_2\, b_{11} - T(R)\, b_{12} \Big);\nonumber\\
&&\hspace*{-5mm} \Delta_1 z_i{}^j = \frac{\alpha}{\pi} C(R)_i{}^j\, g_{11} - \frac{1}{4\pi^2} \lambda^*_{imn}\lambda^{jmn}\, g_{12};\qquad\nonumber\\
&&\hspace*{-5mm} \Delta_1 \lambda^{ijk} = \frac{3\alpha}{2\pi} C(R)_m{}^{(i} \lambda^{jk)m} l_{11} - \frac{3}{8\pi^2} \lambda^*_{mnp} \lambda^{p(ij} \lambda^{k)mn} l_{12} + \frac{\alpha}{4\pi} \lambda^{ijk} \Big(3C_2 d_{11} - T(R) d_{12}\Big),\qquad
\end{eqnarray}

\noindent
where $b_i$, $g_i$, $l_i$, and $d_i$ are finite constants.

The explicit expression for the two-loop anomalous dimension in the HD+MSL scheme can be found in \cite{Kazantsev:2020kfl}. For one-loop finite theories it takes the form

\begin{eqnarray}\label{Gamma_Finite_HD+MSL}
&& (\gamma_{\phi,2})_i{}^j(\alpha,\lambda)\Big|_{\mbox{\scriptsize HD+MSL,\, Eq.} (\ref{1Loop_Finiteness})} = -\frac{3\alpha^2}{2\pi^2} C_2 C(R)_i{}^j \ln \frac{a_\varphi}{a} - \frac{\alpha}{4\pi^2} \Big(\frac{1}{\pi} \lambda{}_{imn}^* \lambda^{jml} C(R)_l{}^n \qquad\nonumber\\
&& + 2\alpha \left[C(R)^2\right]_i{}^j\Big) (A-B).
\end{eqnarray}

\noindent
To construct a two-loop anomalous dimension of the matter superfields for a one-loop finite theory in a general subtraction scheme, we substitute the expressions (\ref{Finite_Renormalization_Explicit}) and (\ref{Gamma_Finite_HD+MSL}) into Eq. (\ref{New_Gamma}). Taking Eq. (\ref{1Loop_Finiteness_Prime}) into account after some transformations we obtain

\begin{eqnarray}\label{Gamma2_Transformed}
&& (\gamma'_\phi)_i{}^j(\alpha',\lambda')\Big|_{\mbox{\scriptsize Eq. (\ref{1Loop_Finiteness_Prime})}} = g^2 \bigg\{-\frac{3\alpha'{}^2}{2\pi^2} C_2 C(R)_i{}^j\,\Big(\ln \frac{a_\varphi}{a} - b_{11} + b_{12} + d_{11} - d_{12}\Big)  \nonumber\\
&& -\frac{\alpha'}{4\pi^2} \Big(\frac{1}{\pi} \lambda'{}_{imn}^* \lambda'{}^{jml} C(R)_l{}^n + 2\alpha' \left[C(R)^2\right]_i{}^j\Big)\Big(A-B-2l_{12}+2l_{11}\Big)\bigg\} + O(g^3).\qquad
\end{eqnarray}

\noindent
This result generalizes the expression derived in \cite{Kazantsev:2020kfl}, which corresponds to the particular case $l_{11} = g_{11}$, $l_{12} = g_{12}$, and $d_{11} = d_{12}=0$.

A new $\beta$-function can be obtained in a similar way starting from Eq. (\ref{Finite_Renormalization_Beta}). Again the terms in Eq. (\ref{Finite_Renormalization_Beta}) containing $\beta_\lambda$ are not essential in the considered (three-loop) approximation. Really, they are of the order $O(g^5)$, because $\partial \alpha'/\partial \lambda \sim O(g^2)$ due to Eq. (\ref{Finite_Renormalizations_Explicit}) and $\beta_\lambda \sim O(g^{5/2})$ due to Eq. (\ref{1Loop_Finiteness_Prime}). Similarly, the derivative $\partial\alpha'/\partial\alpha$ should be replaced by 1, because for one-loop finite theories $\beta \sim O(g^4)$. However, it is necessary to take into account that the one-loop finiteness conditions are imposed on the couplings $\alpha'$ and $\lambda'$, so that the right hand side should be presented as a Taylor series in the deviations $\alpha'-\alpha$ and $\lambda'-\lambda$. Then, using Eq. (\ref{1Loop_Finiteness_Prime}) a three-loop contribution to the new $\beta$-function can be written as

\begin{eqnarray}\label{Beta3_Transformed}
&& \beta'(\alpha',\lambda')\Big|_{\mbox{\scriptsize Eq.} (\ref{1Loop_Finiteness_Prime})} = g^4 \Big( -\frac{\partial \beta_2}{\partial\alpha}\, \Delta_1\alpha - \frac{\partial \beta_2}{\partial\lambda^{mnp}}\, \Delta_1\lambda^{mnp}\nonumber\\
&&\qquad\qquad\qquad\qquad\qquad\qquad - \frac{\partial \beta_2}{\partial\lambda^*_{mnp}}\, \Delta_1\lambda^*_{mnp} + \beta_3\Big)\bigg|_{\alpha\to\alpha';\ \lambda\to\lambda';\ \mbox{\scriptsize Eq.} (\ref{1Loop_Finiteness_Prime})} + O(g^5),\qquad
\end{eqnarray}

\noindent
where in the right hand side the original arguments $\alpha$ and $\lambda$ should be formally replaced by the new couplings $\alpha'$ and $\lambda'$. Note that deriving this equation we took into account that

\begin{equation}
\beta_1(\alpha)\Big|_{\mbox{\scriptsize Eq.} (\ref{1Loop_Finiteness_Prime})} = -\frac{\alpha^2}{2\pi} \Big(3C_2-T(R)\Big)\bigg|_{\mbox{\scriptsize Eq.} (\ref{1Loop_Finiteness_Prime})} = 0.
\end{equation}

Next, we involve the NSVZ equation substituting $\beta_2(\alpha,\lambda)$ and $\beta_3(\alpha,\lambda)$ from Eq. (\ref{Beta23}). Taking into account that the terms proportional to $(\gamma_{\phi,1})_i{}^j$ vanish due to Eq. (\ref{1Loop_Finiteness_Prime}) we obtain

\begin{eqnarray}
&&\hspace*{-7mm} \beta'(\alpha',\lambda')\Big|_{\mbox{\scriptsize Eq.} (\ref{1Loop_Finiteness_Prime})} = -\frac{g^4\alpha'{}^2}{2\pi r} C(R)_i{}^j \Big( -\frac{\partial (\gamma_{\phi,1})_j{}^i}{\partial\alpha}\, \Delta_1\alpha - \frac{\partial (\gamma_{\phi,1})_j{}^i}{\partial\lambda^{mnp}}\, \Delta_1\lambda^{mnp} - \frac{\partial (\gamma_{\phi,1})_j{}^i}{\partial\lambda^*_{mnp}}\, \Delta_1\lambda^*_{mnp} \nonumber\\
&&\hspace*{-7mm} + (\gamma_{\phi,2})_j{}^i \Big)\bigg|_{\alpha\to \alpha';\ \lambda\to\lambda';\ \mbox{\scriptsize Eq.} (\ref{1Loop_Finiteness_Prime})} + O(g^5) = -\frac{g^2\alpha'{}^2}{2\pi r} C(R)_i{}^j (\gamma_\phi')_j{}^i(\alpha',\lambda')\Big|_{\mbox{\scriptsize Eq.} (\ref{1Loop_Finiteness_Prime})} + O(g^5).
\end{eqnarray}

\noindent
Thus, we see that the NSVZ equation really takes place in the first non-trivial approximation independently of the subtraction scheme in a complete agreement with \cite{Kazantsev:2020kfl}. An explicit expression for the function $\beta'$ obtained from this equation is written as

\begin{eqnarray}\label{Beta_Prime_Finite}
&& \beta'(\alpha',\lambda')\Big|_{\mbox{\scriptsize Eq.} (\ref{1Loop_Finiteness_Prime})} = g^4\bigg[\frac{3\alpha'{}^4}{4\pi^3 r} C_2\, \mbox{tr}\left[C(R)^2\right] \Big(\ln \frac{a_\varphi}{a} - b_{11} + b_{12} + d_{11} - d_{12}\Big) + \frac{\alpha'{}^3}{8\pi^3 r}   \nonumber\\
&& \times \Big(\frac{1}{\pi} C(R)_j{}^i C(R)_l{}^n \lambda'{}^*_{imn} \lambda'{}^{jml} + 2\alpha'\,\mbox{tr}\left[C(R)^3\right]\Big) \Big(A-B - 2l_{12} + 2l_{11}\Big)\bigg] + O(g^5). \vphantom{\frac{1}{\pi^2}}\qquad
\end{eqnarray}

\noindent
In the particular case $l_{11}=g_{11}$, $l_{12}=g_{12}$, and $d_{11}=d_{12}=0$ it coincides with the result presented in Ref. \cite{Kazantsev:2020kfl}.

Now, let us consider the Yukawa $\beta$-function and find out whether it satisfies Eq. (\ref{L_Yukawa_Beta_Minimal}) in an arbitrary subtraction scheme. Evidently, this equation is valid in the HD+MSL scheme in all orders, because in this scheme finite constants are not present both in the renormalization constant for matter superfields and in the equation which relates bare and renormalized Yukawa couplings. The first coefficient of the function $(\beta_\lambda)^{ijk}$ is given by the expression

\begin{equation}\label{Beta_Lambda_1Loop}
(\beta_{\lambda,1})^{ijk}  = \frac{3}{2}\, (\gamma_{\phi,1})_m{}^{(i} \lambda^{jk)m} = \frac{3}{2} \Big( - \frac{\alpha}{\pi}C(R)_m{}^{(i}+\frac{1}{4\pi^2}\lambda^*_{mpq}\lambda^{pq(i}\Big) \lambda^{jk)m}.
\end{equation}

\noindent
Certainly, this expression vanishes for one-loop finite theories due to Eq. (\ref{1Loop_Finiteness}). For these theories the first nontrivial (two-loop) contribution to the Yukawa $\beta$-function in the HD+MSL scheme is written as

\begin{eqnarray}\label{Beta_Lambda_2Loop}
&&\hspace*{-7mm} (\beta_{\lambda,2})^{ijk}\Big|_{\mbox{\scriptsize HD+MSL,\, Eq.} (\ref{1Loop_Finiteness})}  = \frac{3}{2}\, (\gamma_{\phi,2})_m{}^{(i} \lambda^{jk)m}\Big|_{\mbox{\scriptsize HD+MSL,\, Eq.} (\ref{1Loop_Finiteness})} = - \frac{9\alpha^2}{4\pi^2}\, \ln \frac{a_\varphi}{a} C_2 C(R)_m{}^{(i} \lambda^{jk)m}\nonumber\\
&&\hspace*{-7mm} - \frac{3\alpha}{8\pi^2} (A-B) \Big(\frac{1}{\pi} \lambda{}_{mnp}^* \lambda^{lp(i} C(R)_l{}^n  + 2\alpha \left[C(R)^2\right]_m{}^{(i} \Big) \lambda^{jk)m}.
\end{eqnarray}

\noindent
Next, we perform the finite renormalization (\ref{Finite_Renormalization_Explicit}) and impose the one-loop finiteness conditions (\ref{1Loop_Finiteness_Prime}) on the new couplings $\alpha'$ and $\lambda'$. A two-loop contribution to the new Yukawa $\beta$-function can be found from Eq. (\ref{Finite_Renormalization_Beta_Lambda}). In this approximation the only essential term is the one containing the derivative $\partial\lambda'/\partial\lambda$, but, as earlier, one should take into account the difference between the conditions (\ref{1Loop_Finiteness}) and (\ref{1Loop_Finiteness_Prime}). Then, making the calculations similar to those for the functions $\beta$ and $(\gamma_\phi)_i{}^j$, in the first nontrivial approximation we obtain

\begin{eqnarray}\label{New_Beta_Lambda}
&&\hspace*{-3mm} (\beta_\lambda')^{ijk}(\alpha',\lambda')\Big|_{\mbox{\scriptsize Eq.} (\ref{1Loop_Finiteness_Prime})} = g^{5/2}\Big( - \frac{\partial (\beta_{\lambda,1})^{ijk}}{\partial\alpha} \Delta_1\alpha - \frac{\partial (\beta_{\lambda,1})^{ijk}}{\partial\lambda^{mnp}} \Delta_1\lambda^{mnp} - \frac{\partial (\beta_{\lambda,1})^{ijk}}{\partial\lambda^*_{mnp}} \Delta_1\lambda^*_{mnp}\nonumber\\
&&\hspace*{-3mm} + (\beta_{\lambda,2})^{ijk} \Big)\Big|_{\alpha\to\alpha';\ \lambda\to\lambda';\ \mbox{\scriptsize Eq.} (\ref{1Loop_Finiteness_Prime})} + O(g^{7/2}).
\end{eqnarray}

\noindent
The expressions entering this equation are given by Eqs. (\ref{Finite_Renormalization_Explicit}), (\ref{Beta_Lambda_1Loop}), and (\ref{Beta_Lambda_2Loop}). Substituting them in Eq. (\ref{New_Beta_Lambda}) we find the explicit expression for the transformed Yukawa $\beta$-function,

\begin{eqnarray}
&&\hspace*{-8mm} (\beta_{\lambda,2}')^{ijk}(\alpha',\lambda')\Big|_{\mbox{\scriptsize Eq.} (\ref{1Loop_Finiteness_Prime})} = \bigg\{\, \frac{9\alpha'{}^2}{4\pi^2} C_2 C(R)_m{}^{(i} \lambda'{}^{jk)m} \Big(-\ln \frac{a_\varphi}{a} +b_{11}- b_{12}-d_{11}+d_{12}\Big) - \frac{3\alpha'}{8\pi^2}\nonumber\\
&&\hspace*{-8mm} \times \Big(\frac{1}{\pi} \lambda'{}^*_{mnp} \lambda'{}^{lp(i} C(R)_l{}^n  + 2\alpha' \left[C(R)^2\right]_m{}^{(i} \Big) \lambda'{}^{jk)m} \Big(A-B + 2l_{11} - 2 l_{12}\Big) \bigg\}.
\end{eqnarray}

\noindent
Comparing this result with the expression for the transformed anomalous dimension given by Eq. (\ref{Gamma2_Transformed}) we conclude that Eq. (\ref{L_Yukawa_Beta_Minimal}) with $L=2$ for one-loop finite theories is really satisfied in an arbitrary subtraction scheme,

\begin{equation}
(\beta_{\lambda,2}')^{ijk}(\alpha',\lambda')\Big|_{\mbox{\scriptsize Eq.} (\ref{1Loop_Finiteness_Prime})} = \frac{3}{2}\, (\gamma_{\phi,2})_m{}^{(i}(\alpha',\lambda') \lambda'{}^{jk)m}\Big|_{\mbox{\scriptsize Eq.} (\ref{1Loop_Finiteness_Prime})},
\end{equation}

\noindent
although both sides of this equation are scheme dependent.

Note that for theories which are not finite in the one-loop approximation Eq. (\ref{L_Yukawa_Beta_Minimal}) with $L=2$ does not hold after the finite renormalization (\ref{Finite_Renormalization_Explicit}). Really, in this case after some transformations we obtain

\begin{eqnarray}
&& g^{1/2} \lambda'{}^{ijk} = g^{1/2} \lambda_0^{mnp} (Z_\phi'{}^{1/2})_m{}^i (Z_\phi'{}^{1/2})_n{}^j (Z_\phi'{}^{1/2})_p{}^k  - g^{3/2}\bigg\{\frac{3\alpha}{2\pi} C(R)_m{}^{(i} \lambda^{jk)m} (g_{11} - l_{11})\qquad \nonumber\\
&& - \frac{3}{8\pi^2} \lambda^*_{mnp} \lambda^{np(i} \lambda^{jk)m} (g_{12} - l_{12}) - \frac{\alpha}{4\pi}\lambda^{ijk}\Big(3C_2 d_{11} -  T(R) d_{12}\Big)\bigg\} + O(g^{5/2}).
\end{eqnarray}

\noindent
After differentiating with respect to $\ln\mu$, the first term in the right hand side produces the right hand side of Eq. (\ref{L_Yukawa_Beta_Minimal}) with $L=2$ for the new couplings, while the remaining ones give nontrivial corrections to this equation proportional to $g^{5/2}$ if one-loop RGFs do not vanish. Therefore, after the finite renormalization (\ref{Finite_Renormalization_Explicit})

\begin{equation}
(\beta_{\lambda,2}')^{ijk}(\alpha',\lambda') = \frac{3}{2}\, (\gamma_{\phi,2})_m{}^{(i}(\alpha',\lambda') \lambda'{}^{jk)m} + \mbox{terms proportional to $\beta_1$ and $\gamma_{\phi,1}$}.
\end{equation}

\noindent
Consequently, Eq. (\ref{Yukawa_Beta_Minimal}) in the two-loop approximation can be broken if the theory is not finite in the one-loop approximation. However, for one-loop finite theories Eq. (\ref{L_Yukawa_Beta_Minimal}) with $L=2$ is always valid, because in this case $\beta_1=0$ and $(\gamma_{\phi,1})_i{}^j = 0$.

\section{Finite renormalizations which do not break Eq. (\ref{Assumption}) for $L=3$}
\hspace*{\parindent}\label{Section_Finiteness_For_L=3}

Now, let us construct finite renormalizations which do not break finiteness in the next ($L=3$) approximation. As we discussed above, in this approximation (for one-loop finite theories) the exact expressions for the gauge and Yukawa $\beta$-functions are valid for an arbitrary renormalization prescription. Therefore, it is sufficient to find finite renormalizations under that the condition

\begin{equation}
(\gamma_{\phi,2})_i{}^j(\alpha,\lambda) = 0
\end{equation}

\noindent
remains invariant. Evidently, the expression (\ref{Gamma_Finite_HD+MSL}) vanishes if $a_\varphi = a$ and $A=B$. In this case the anomalous dimension (\ref{Gamma2_Transformed}) is equal to 0 if the parameters of the finite renormalization satisfy the constraints

\begin{equation}
b_{11} - d_{11} = b_{12} - d_{12};\qquad l_{11} = l_{12}.
\end{equation}

\noindent
Substituting them into Eq. (\ref{Finite_Renormalization_Explicit}) we see that under the considered finite renormalizations for theories finite in the one-loop approximation the couplings change as

\begin{equation}\label{Couplings_Shift}
\Delta_1\alpha\Big|_{\mbox{\scriptsize Eq.} (\ref{1Loop_Finiteness})} = \frac{3\alpha^2 C_2}{2\pi} (b_{11} - b_{12});\qquad \Delta_1\lambda^{ijk}\Big|_{\mbox{\scriptsize Eq.} (\ref{1Loop_Finiteness})} = \frac{3\alpha}{4\pi} \lambda^{ijk} C_2 (b_{11}- b_{12}),
\end{equation}

\noindent
while

\begin{equation}
\Delta_1z_i{}^j = \frac{\alpha}{\pi} C(R)_i{}^j (g_{11}-g_{12})
\end{equation}

\noindent
contains the parameters $g_{11}$ and $g_{12}$ which can take arbitrary values.

In general, the expressions (\ref{Couplings_Shift}) do not vanish. However, it is possible to construct a function of the couplings that remains invariant under these transformations in the lowest nontrivial approximation. Really, if we consider the expression (\ref{P_Definition}) (depending on $\alpha$ and $\lambda$) and impose the one-loop finiteness conditions (\ref{1Loop_Finiteness_Prime}) (for the transformed couplings $\alpha'$ and $\lambda'$), then from Eq. (\ref{Couplings_Shift}) we obtain

\begin{eqnarray}\label{P_Condition_Lowest}
&&\hspace*{-3mm} g P_i{}^j(\alpha,\lambda)\Big|_{\mbox{\scriptsize Eq.} (\ref{1Loop_Finiteness_Prime})} = - g \Big(P_i{}^j(\alpha',\lambda') - P_i{}^j(\alpha,\lambda) \Big)\Big|_{\mbox{\scriptsize Eq.} (\ref{1Loop_Finiteness_Prime})}\nonumber\\
&&\hspace*{-3mm}\quad = - g^2\Big(2\Delta_1\lambda^*_{imn} \lambda^{jmn} + 2\lambda^*_{imn}\Delta_1 \lambda^{jmn} - 8\pi \Delta_1\alpha C(R)_i{}^j\Big)\Big|_{\mbox{\scriptsize Eq.} (\ref{1Loop_Finiteness_Prime})} + O(g^3)  = O(g^3).\qquad\vphantom{\frac{1}{2}}
\end{eqnarray}

\noindent
This implies that in the considered approximation this expression does not change under the finite renormalizations that we are interested in. In the next section we will see that a generalization of this statement can be used for constructing finite renormalizations which do not break Eq. (\ref{Assumption}) for an arbitrary value of $L$.

\section{The result for an arbitrary $L$}
\hspace*{\parindent}\label{Section_Main_Result}

Now, let us assume that the renormalization scheme is tuned in such a way that Eq. (\ref{Assumption}) is satisfied. Our purpose is to prove that Eqs. (\ref{L_Yukawa_Beta_Minimal}) and (\ref{L_NSVZ}) are satisfied for an arbitrary renormalization prescription which respects finiteness in the previous loops. First, we note that at least one such subtraction scheme exists, because, according to \cite{Kazakov:1986bs,Ermushev:1986cu,Lucchesi:1987he,Lucchesi:1987ef}, there is a renormalization prescription for which a one-loop finite theory is finite in all loops. Evidently, in this case Eqs. (\ref{L_Yukawa_Beta_Minimal}) and (\ref{L_NSVZ}) are valid. Therefore, we may start with a subtraction scheme in which these equations are satisfied. (RGFs and couplings in this scheme will be denoted by letters without primes.) The other renormalization prescriptions are related to this scheme by finite renormalizations (\ref{Finite_Renormalizations}). Taking Eq. (\ref{Finite_Renormalizations_Explicit}) into account we see that after the replacement (\ref{Replacement})

\begin{eqnarray}
&& \frac{\partial\alpha'}{\partial\alpha} = 1 + O(g); \qquad \frac{\partial\alpha'}{\partial \lambda^{ijk}} = O(g^{2}); \qquad\ \frac{\partial\ln z_i{}^j}{\partial\alpha} = O(g);\qquad \frac{\partial\ln z_i{}^j}{\partial\lambda^{mnp}} = O(g);\qquad\nonumber\\
&&\qquad \frac{\partial\lambda'{}^{ijk}}{\partial\alpha} = O(g);\qquad\quad \frac{\partial\lambda'{}^{ijk}}{\partial \lambda^{mnp}} = \delta^{(i}_m \delta^j_n \delta^{k)}_p + O(g);\qquad\quad \frac{\partial\lambda'{}^{ijk}}{\partial\lambda^*_{mnp}} = O(g).
\end{eqnarray}

\noindent
Moreover, due to Eq. (\ref{Assumption})

\begin{equation}
\beta = O(g^{L+2}); \qquad\quad\ (\gamma_\phi)_i{}^j = O(g^L);\qquad\  (\beta_\lambda)^{ijk} = O(g^{L+1/2}).
\end{equation}

\noindent
Substituting the above equations into Eqs. (\ref{Finite_Renormalization_Beta}), (\ref{Finite_Renormalization_Gamma}), and (\ref{Finite_Renormalization_Beta_Lambda}) we obtain that after the considered finite renormalization

\begin{eqnarray}\label{New_RGFs}
&& \beta'(\alpha',\lambda')\Big|_{\mbox{\scriptsize Eq.} (\ref{1Loop_Finiteness_Prime})} = \beta(\alpha,\lambda)\Big|_{\mbox{\scriptsize Eq.} (\ref{1Loop_Finiteness_Prime})} + O(g^{L+3});\qquad \nonumber\\
&& (\gamma_\phi')_i{}^j(\alpha',\lambda')\Big|_{\mbox{\scriptsize Eq.} (\ref{1Loop_Finiteness_Prime})} = (\gamma_\phi)_i{}^j(\alpha,\lambda)\Big|_{\mbox{\scriptsize Eq.} (\ref{1Loop_Finiteness_Prime})} + O(g^{L+1}); \qquad\nonumber\\
&& (\beta_\lambda')^{ijk}(\alpha',\lambda')\Big|_{\mbox{\scriptsize Eq.} (\ref{1Loop_Finiteness_Prime})} = (\beta_\lambda)^{ijk}(\alpha,\lambda)\Big|_{\mbox{\scriptsize Eq.} (\ref{1Loop_Finiteness_Prime})} + O(g^{L+3/2}).
\end{eqnarray}

\noindent
Note that according to our assumption, Eqs. (\ref{L_Yukawa_Beta_Minimal}) and (\ref{L_NSVZ}) for RGFs without primes are valid under the condition (\ref{1Loop_Finiteness}), which is different from the condition (\ref{1Loop_Finiteness_Prime}).

First, let us construct finite renormalizations which do not break Eq. (\ref{Assumption}). Earlier it was demonstrated that in the particular case $L=3$ such finite renormalizations satisfy Eq. (\ref{P_Condition_Lowest}). For an arbitrary value of $L$ we will prove that they are determined by the condition

\begin{equation}\label{P_Condition}
g P_i{}^j(\alpha,\lambda)\Big|_{\mbox{\scriptsize Eq.} (\ref{1Loop_Finiteness_Prime})} = O(g^L),
\end{equation}

\noindent
which is a generalization of Eq. (\ref{P_Condition_Lowest}). This can be done with the help of mathematical induction. For $L=2$ this equation is trivial, and for $L=3$ it has been verified in Sect. \ref{Section_Finiteness_For_L=3}. Let us suppose that, under the assumption (\ref{Assumption}) with $L=k$, Eq. (\ref{P_Condition}) is valid for $L=k$ (where $k\ge 2$) and prove that, under the assumption (\ref{Assumption}) with $L=k+1$, it is also valid for $L=k+1$.

Let Eq. (\ref{Assumption}) be valid for $L=k+1$ for both original and new couplings. Evidently, finite renormalizations relating these couplings should be searched among such finite renormalizations that do not break Eq. (\ref{Assumption}) with $L=k$. In the latter case RGFs satisfy Eq. (\ref{New_RGFs}) with $L=k$. Therefore, the finite renormalizations that we are interested in can be found from the condition

\begin{equation}\label{New_Gamma_Induction}
O(g^{k+1}) = (\gamma_\phi')_i{}^j(\alpha',\lambda')\Big|_{\mbox{\scriptsize Eq.} (\ref{1Loop_Finiteness_Prime})} = (\gamma_\phi)_i{}^j(\alpha,\lambda)\Big|_{\mbox{\scriptsize Eq.} (\ref{1Loop_Finiteness_Prime})} + O(g^{k+1}).
\end{equation}

\noindent
If we impose only the constraint $T(R) = 3C_2$ (which does not depend on couplings), then the anomalous dimension of the matter superfields will differ from the one for a one-loop finite theory by terms proportional to various tensor structures containing powers of $g P$. We will schematically denote these structures by $(c(g)\cdot gP \cdot \ldots \cdot gP)_i{}^j$, where the tensor $c(g)$ contains couplings, group Casimirs, and numerical factors which cannot be combined into $gP$. Using this notation the anomalous dimension of the matter superfields for a theory with $T(R) = 3C_2$ can be written as

\begin{equation}\label{P_Expansion_Gamma_N}
(\gamma_\phi)_i{}^j(\alpha,\lambda)\Big|_{T(R)=3C_2} = (\gamma_\phi)_i{}^j(\alpha,\lambda)\Big|_{\mbox{\scriptsize Eq.} (\ref{1Loop_Finiteness})} + (c_1(g)\cdot gP + c_2(g)\cdot gP\cdot gP + \ldots )_i{}^j
\end{equation}

\noindent
It is important that the coefficient of the lowest order term containing $gP$ does not vanish due to Eq. (\ref{1Loop_Gamma}),

\begin{equation}\label{First_Coefficient}
(c_1(0)\cdot g P)_i{}^j = \frac{g}{8\pi^2} P_i{}^j(\alpha,\lambda) \ne 0.
\end{equation}

\noindent
Let us consider the expression (\ref{P_Expansion_Gamma_N}) for such values of couplings that $\lambda'{}^*_{imn} \lambda'{}^{jmn} = 4\pi\alpha' C(R)_i{}^j$. In other words, we impose the constraint (\ref{1Loop_Finiteness_Prime}). Evidently, if the condition (\ref{Assumption}) is valid for $L=k+1$, then it also holds for $L=k$. This implies that we can use the induction hypothesis, according to which, the expression $g P_i{}^j$ is of the order $O(g^k)$ and can presented in the form

\begin{equation}\label{P_Structure}
g P_i{}^j(\alpha,\lambda)\Big|_{\mbox{\scriptsize Eq.} (\ref{1Loop_Finiteness_Prime})} = g^k (p_k)_i{}^j + O(g^{k+1}).
\end{equation}

\noindent
Therefore, terms containing the $m$-th power of $gP$ in Eq. (\ref{P_Expansion_Gamma_N}) are proportional to $g^{mk}$. Taking into account that $k\ge 2$, we see that $mk<k+1$ only for $m=1$. Consequently, if we consider only terms proportional to $g^k$, then it is possible to ignore all terms with $m>1$ and keep only the lowest term linear in $g P$,

\begin{eqnarray}
&& (\gamma_\phi)_i{}^j(\alpha,\lambda)\Big|_{\mbox{\scriptsize Eq.} (\ref{1Loop_Finiteness_Prime})} =  (\gamma_\phi)_i{}^j(\alpha,\lambda)\Big|_{\mbox{\scriptsize Eq.} (\ref{1Loop_Finiteness})} + g^k (c_1(0) \cdot p_k)_i{}^j + O(g^{k+1})\nonumber\\
&& \qquad\qquad\qquad\qquad\qquad\qquad\qquad\qquad =  (\gamma_\phi)_i{}^j(\alpha,\lambda)\Big|_{\mbox{\scriptsize Eq.} (\ref{1Loop_Finiteness})} + \frac{g^k}{8\pi^2} (p_k)_i{}^j + O(g^{k+1}),\qquad
\end{eqnarray}

\noindent
where we also used Eq. (\ref{First_Coefficient}).  According to Eq. (\ref{New_Gamma_Induction}), the left hand side of this equation is of the order $O(g^{k+1})$. From the other side, according to the assumption (\ref{Assumption}) for $L=k+1$, the anomalous dimension in the right hand side is of the order $O(g^{k+1})$. Therefore, comparing the terms proportional to $g^k$ in the left and right hand sides we obtain

\begin{equation}
(p_k)_i{}^j = 0.
\end{equation}

\noindent
Then from Eq. (\ref{P_Structure}) we conclude that Eq. (\ref{P_Condition}) is also satisfied for $L=k+1$. Thus, the inductive step is completed and Eq. (\ref{P_Condition}) is proved.

Next, let us return to Eq. (\ref{New_RGFs}) and consider, e.g., the anomalous dimension of the matter superfields. From Eqs. (\ref{P_Condition}), (\ref{P_Expansion_Gamma_N}), and (\ref{First_Coefficient}) we see that

\begin{eqnarray}\label{Gamma_New}
&& (\gamma_\phi')_i{}^j(\alpha',\lambda')\Big|_{\mbox{\scriptsize Eq.} (\ref{1Loop_Finiteness_Prime})}
= (\gamma_\phi)_i{}^j(\alpha,\lambda)\Big|_{\mbox{\scriptsize Eq.} (\ref{1Loop_Finiteness})} + \frac{g}{8\pi^2} P_i{}^j\Big|_{\mbox{\scriptsize Eq.} (\ref{1Loop_Finiteness_Prime})} + O(g^{L+1})\nonumber\\
&& \qquad\qquad\qquad\qquad\qquad\qquad = (\gamma_\phi)_i{}^j(\alpha,\lambda)\Big|_{\mbox{\scriptsize Eq.} (\ref{1Loop_Finiteness})} + g (\gamma_{\phi,1})_i{}^j(\alpha,\lambda)\Big|_{\mbox{\scriptsize Eq.} (\ref{1Loop_Finiteness_Prime})} + O(g^{L+1}),\qquad
\end{eqnarray}

\noindent
where we took into account that both explicitly written terms in the right hand side are of the order $O(g^L)$, while the other contributions contain larger degrees of $g$. Say, the terms containing higher degrees of $g P$ are proportional at least to $O(g^{2L})$. (Note that deriving Eq. (\ref{Gamma_New}) we also used Eqs. (\ref{1Loop_Gamma}) and (\ref{First_Coefficient}).) Similar equations for the gauge and Yukawa $\beta$-functions can be written as

\begin{eqnarray}\label{Beta_New}
&& \beta'(\alpha',\lambda')\Big|_{\mbox{\scriptsize Eq.} (\ref{1Loop_Finiteness_Prime})} = \beta(\alpha,\lambda)\Big|_{\mbox{\scriptsize Eq.} (\ref{1Loop_Finiteness})} + g^3 \beta_2(\alpha,\lambda)\Big|_{\mbox{\scriptsize Eq.} (\ref{1Loop_Finiteness_Prime})} + O(g^{L+3});\\
\label{Beta_Lambda_New}
&& (\beta_\lambda')^{ijk}(\alpha',\lambda')\Big|_{\mbox{\scriptsize Eq.} (\ref{1Loop_Finiteness_Prime})} = (\beta_\lambda)^{ijk}(\alpha,\lambda)\Big|_{\mbox{\scriptsize Eq.} (\ref{1Loop_Finiteness})} + g^{3/2} (\beta_{\lambda,1})^{ijk}(\alpha,\lambda)\Big|_{\mbox{\scriptsize Eq.} (\ref{1Loop_Finiteness_Prime})} + O(g^{L+3/2}).\qquad
\end{eqnarray}

Let us recall that we started with a renormalization scheme (say, HD+MSL) in which Eqs. (\ref{L_Yukawa_Beta_Minimal}) and (\ref{L_NSVZ}) are satisfied. This implies that under the assumption (\ref{Assumption}) the first nontrivial coefficients of RGFs in this scheme are related by the equations

\begin{eqnarray}\label{Beta_L_Equations}
&& \beta_{L+1}(\alpha,\lambda)\Big|_{\mbox{\scriptsize Eq.} (\ref{1Loop_Finiteness})} = - \frac{\alpha^2}{2\pi r} C(R)_i{}^j (\gamma_{\phi,L})_j{}^i(\alpha,\lambda)\Big|_{\mbox{\scriptsize Eq.} (\ref{1Loop_Finiteness})};\qquad \nonumber\\
&& (\beta_{\lambda,L})^{ijk}(\alpha,\lambda)\Big|_{\mbox{\scriptsize Eq.} (\ref{1Loop_Finiteness})} = \frac{3}{2}\, (\gamma_{\phi,L})_m{}^{(i}(\alpha,\lambda) \lambda^{jk)m}\Big|_{\mbox{\scriptsize Eq.} (\ref{1Loop_Finiteness})}.
\end{eqnarray}

\noindent
Also we note that the two-loop gauge $\beta$-function, the one-loop anomalous dimension of the matter superfield, and the one-loop Yukawa $\beta$-function are scheme independent and satisfy Eqs. (\ref{L_Yukawa_Beta_Minimal}) and (\ref{L_NSVZ}) for any renormalization prescription (without the use of the one-loop finiteness conditions). This implies that the equations

\begin{equation}\label{Beta_Lowest_Equations}
\beta_{2}(\alpha,\lambda) = - \frac{\alpha^2}{2\pi r} C(R)_i{}^j (\gamma_{\phi,1})_j{}^i(\alpha,\lambda);\qquad
(\beta_{\lambda,1})^{ijk}(\alpha,\lambda) = \frac{3}{2}\, (\gamma_{\phi,1})_m{}^{(i}(\alpha,\lambda) \lambda^{jk)m}
\end{equation}

\noindent
are valid for arbitrary values of $\alpha$ and $\lambda^{ijk}$.

Two explicitly written terms in the right hand side of Eq. (\ref{Gamma_New}) are of the order $O(g^L)$. Similarly, explicitly written terms in the right hand sides of Eqs. (\ref{Beta_New}) and (\ref{Beta_Lambda_New}) are of the order $O(g^{L+2})$ and $O(g^{L+1/2})$, respectively. Therefore, from the above equations we obtain

\begin{eqnarray}\label{Beta_Result}
&& \beta'(\alpha',\lambda')\Big|_{\mbox{\scriptsize Eq.} (\ref{1Loop_Finiteness_Prime})} = - \frac{g^2 \alpha'{}^2}{2\pi r} C(R)_i{}^j (\gamma_{\phi}')_j{}^i(\alpha',\lambda')\Big|_{\mbox{\scriptsize Eq.} (\ref{1Loop_Finiteness_Prime})} + O(g^{L+3});\\
\label{Beta_Lambda_Result}
&& (\beta_\lambda')^{ijk}(\alpha',\lambda')\Big|_{\mbox{\scriptsize Eq.} (\ref{1Loop_Finiteness_Prime})} = \frac{3}{2}\, g^{1/2} (\gamma_{\phi}')_m{}^{(i}(\alpha',\lambda') \lambda'{}^{jk)m}\Big|_{\mbox{\scriptsize Eq.} (\ref{1Loop_Finiteness_Prime})} + O(g^{L+3/2}).
\end{eqnarray}

\noindent
Note that we also took into account that the differences $\alpha'-\alpha$ and $\lambda'-\lambda$ are proportional to the parameter $g$ and give terms unessential in the considered approximation. From Eqs. (\ref{Beta_Result}) and (\ref{Beta_Lambda_Result}) we conclude that Eqs. (\ref{L_Yukawa_Beta_Minimal}) and (\ref{L_NSVZ}) are valid for an arbitrary renormalization prescription consistent with Eq. (\ref{Assumption}).

Thus, for a one-loop finite theory in an arbitrary subtraction scheme for which Eq. (\ref{Assumption}) is satisfied the $L$-loop Yukawa and $(L+1)$-loop gauge $\beta$-functions are related to the $L$-loop anomalous dimension of the matter superfields. This statement generalizes the observation made in \cite{Kazantsev:2020kfl} that for one-loop finite theories the three-loop gauge $\beta$-function is related to the two-loop anomalous dimension independently of a subtraction scheme, see Sect. \ref{Section_Three_Loops}.

As a simple consequence we obtain the result of Ref. \cite{Grisaru:1985tc} that for a theory finite in $L$-loops the $(L+1)$-loop $\beta$-function vanishes. Really, for such a theory $(\gamma_{\phi,L})_i{}^j(\alpha,\lambda) = 0$. Therefore, in this case from Eq. (\ref{L_NSVZ}) we conclude that $\beta_{L+1}(\alpha,\lambda) = 0$. (Originally, in Ref. \cite{Grisaru:1985tc}, this statement was derived by a different method.)

\section{Conclusion}
\hspace*{\parindent}

We considered a theory finite in the one-loop approximation assuming that a subtraction scheme is tuned in such a way that the first nontrivial contribution to the gauge $\beta$-function comes from the $(L+1)$-th loop, while the first nontrivial contributions to the Yukawa $\beta$-function and to the anomalous dimension of the matter superfields appear only in the $L$-th loop. It was demonstrated that in this case {\it for any renormalization prescription} which does not break the above conditions the $(L+1)$-loop gauge $\beta$-function and the $L$-loop Yukawa $\beta$-function are related to the $L$-loop anomalous dimension by simple algebraic equations. In other words, under the above formulated assumption Eq. (\ref{Yukawa_Beta_Minimal}) and the NSVZ equation (\ref{NSVZ_Beta1}) are valid in the first non-trivial order independently of a subtraction scheme, although both sides of these equations are scheme-dependent. (In the particular case $L=2$ this statement exactly agrees with the explicit calculations made in \cite{Kazantsev:2020kfl}.) As a straightforward consequence, we obtain a simple proof of the theorem (first derived in \cite{Grisaru:1985tc}) that for a theory finite in a certain order the $\beta$-function vanishes in the next order.

\vspace{5mm}

\section*{Acknowledgments}
\hspace*{\parindent}

The author would like to express the gratitude to A.E.Kazantsev for valuable discussions.

This research has been supported by the Foundation for the Advancement of Theoretical Physics and Mathematics `BASIS', grant No. 19-1-1-45-1.

\end{document}